\documentclass[3p,times,twocolumn]{elsarticle}

\usepackage{lineno}
\usepackage{lineno,hyperref} 

\journal{Computers and Electronics in Agriculture}

\usepackage{multirow}
\usepackage{float}
\usepackage{subfloat} 
 \usepackage[caption]{subfig}
\usepackage{url}
\usepackage{hyperref}

\usepackage[T1]{fontenc}
\usepackage[utf8]{inputenc}
\usepackage[font=small]{caption}
\usepackage[export]{adjustbox}
\usepackage{array}

\begin{document}

\begin{frontmatter}

\title{Wireless Technologies for Agricultural Monitoring using Internet of Things Devices with Energy Harvesting Capabilities}

\author[mymainaddress]{Sebastian Sadowski}

\author[mymainaddress]{Petros Spachos\corref{mycorrespondingauthor}}
\cortext[mycorrespondingauthor]{Corresponding author}
\ead{petros@uoguelph.ca}

\address[mymainaddress]{School of Engineering, University of Guelph, Guelph, Ontario, N1G 2W1, Canada}

\begin{abstract}  
Technological advances in the Internet of Things (IoT) have lead the way for technology to be used in ways that were never possible before. Through the development of devices with low-power radios, Wireless Sensor Networks (WSN) can be configured for almost any type of application. Agricultural has been one example where IoT and WSN have been able to increase productivity, efficiency, and output yield. Systems that previously required manual operation can be easily replaced with sensors and actuators to automate the process such as irrigation and disease management. Powering these devices is a concern as batteries are often required due to devices being located where electricity is not readily available. In this paper, a comparison is performed between three wireless technologies: IEEE 802.15.4 (Zigbee), Long Range Wireless Area Network (LoRaWAN), and IEEE 802.11g (WiFi 2.4~GHz) for agricultural monitoring with energy harvesting capabilities. According to experimental results, LoRaWAN is the optimal technology to use in an agricultural monitoring system where power consumption and network lifetime are a priority. The experimental results can be used for the selection of wireless technology for agricultural monitoring following application requirements.
\end{abstract}


\end{frontmatter}


\section{Introduction}

In the era of Internet of Things (IoT), everyday objects are equipped with microcontrollers and communication devices that work together to improve one's quality of life~\cite{iot, iot2}. While IoT has been of great value to society in the automation of tasks, IoT devices often consume a large amount of energy~\cite{iotEnergy, tongke2013smart}. Energy consumption comes from  various processes such as sensing systems, application operating systems, and the communication radio~\cite{iotPower}. In order to improve energy efficiency, each of the individual processes needs to be optimized~\cite{agriculturalproduction}. When it comes to IoT devices, processes such as the operating and sensing systems are often based on the application requirements and difficult to reduce. The easiest function to modify and optimize the device power consumption is the communication radio. 

IoT devices can be used in monitoring systems consisting of nodes that interact with the environment using sensors to gather real-time information and transmit it to a destination. In every monitoring system~\cite{wsn}, power consumption is often a top concern in order for the system to function properly. If a sensor node ceases to transmit, information at the node's location would be missing and the system would no longer behave accurately. In order to optimize power consumption, the application requirements are required. One application where IoT monitoring systems can be used is in agricultural. When IoT and Wireless Sensor Networks (WSN) are used in agricultural, advanced farming techniques can be applied which is known as Precision Agricultural (PA)~\cite{precisionFarming}. 

PA allows for a greater amount of control in the growing of crops and the raising of livestock. By using technology to monitor crops, the efficiency can be increased and costs can be reduced since more precise remedies can be applied to crops~\cite{iotagriculture}. In PA applications WSN often consist of batteries that are used to power the sensor nodes while outside. This is a major issue as after the battery dies either the battery will need to be replaced or if possible, recharged. Due to the node being outdoors, rechargeable batteries and an energy harvesting device can be used. Since solar power is readily available it can be easily harvested in-order to allow for the sensor node to function for a longer period of time. One of the most optimal methods of optimizing the battery life of nodes in WSN is through the wireless technology communicating the node's information. In agricultural applications the most commonly used wireless technologies have been: IEEE 802.15.4 (Zigbee) \cite{zigbee}, Long Range Wireless Area Network (LoRaWAN) \cite{lorawan}, and IEEE 802.11 (WiFi) \cite{wifijournal} \cite{agriculturalTechnologies}. By using a different technology the lifetime of the nodes could be increased, in-turn extending the lifetime of the network. 

Batteries have become a major source of energy used in a variety of electrical components and devices. Disposable single-use batteries are often used in portable devices. Once the energy in the battery has been depleted, it must be properly disposed and replaced in the device. In recent years, rechargeable batteries have become popular due to their ability to be recharged and reused multiple times, reducing the amount of waste created. Through the development of devices with efficient energy harvesting capabilities, rechargeable batteries have become more efficient. Devices are now able to be scattered in remote locations and function for long periods of time through the use of batteries that can quickly collect energy and recharged. 

One application that can greatly benefit from the use of rechargeable batteries and energy harvesting is agriculture. New technologies and IoT devices have revolutionized the way farmers are able to interact and monitor their growth. By combining traditional approaches, such as energy harvesting techniques, with  IoT devices, PA can be performed.  A promising solution toward achieving PA is the use of IoT system with energy harvesting capabilities. The IoT uses small, low-power embedded electronics that transmit data across a network.   Often, when sensor nodes are configured and placed outdoors in a field, a power source is required. However, in a field, electricity is not readily available and batteries must be used. Due to the need to replace batteries once depleted, rechargeable-batteries are often used.

In this paper, through extensive experimentation, a comparison between the power consumption of three wireless technologies: Zigbee, LoRaWAN, and IEEE 802.11g (WiFi 2.4 GHz) is performed. The technologies were selected based on the prevalence and popularity in agricultural applications. They are compared through the use of an agricultural monitoring system using IoT devices with solar energy harvesting capabilities. Three identical systems were created each performing identical tasks functioning using different wireless technologies.
 
The main contributions of this work are as follows:
\begin{itemize}
    \item A prototype was designed using IoT components for agricultural monitoring applications with solar energy harvesting capabilities. 
    
    \item Extensive experimentation was performed demonstrating the benefits of using IoT devices with energy harvesting for agricultural applications. In total, three experiments were conducted throughout a year and in different seasons.
     
\item The experimental results can be used as an indicator of the power requirements of the wireless technologies when used for agricultural monitoring with IoT devices.
    
    \item According to experimental results, LoRaWAN is the optimal wireless technology to use for agricultural monitoring if power consumption and network lifetime are a concern.
     
\end{itemize}

The rest of this paper is organized as follows: Section~\ref{related} reviews the related work on wireless technologies in agricultural applications. In Section~\ref{frame}, a framework of the designed system is presented, followed by Section~\ref{setup}, with a description of the experimental procedure. The experimental results and a discussion are presented in Section~\ref{results}. Finally, Section~\ref{con} concludes this work.

\section{Related Work} \label{related}
Over the years, applications using IoT devices and WSN in PA have become more commonplace in today's society. By using battery powered sensor nodes combined with traditional farming practices an increase in output efficiency and a reduction in costs can be achieved. While some researchers have focused on implementing energy harvesting to extend sensor node lifetime, others have modified the sensor nodes to use less energy through its standard operating procedure~\cite{agriculture}. 

In~\cite{topology}, a survey was performed studying the lifetime of WSN and the energy saved with different types of network topologies. Based on determined results, there are a lot of problems and issues when selecting a topology for extending the network lifetime. One issue is the trade-offs that need to occur in the network. In order for the network lifetime to be extended trade-offs must occur and other parameters are required to be sacrificed. It was suggested that energy efficient articles be developed to optimize the energy supply. 

In the works~\cite{nitratesensor} and~\cite{watersaving}, systems were proposed for agricultural monitoring. In~\cite{nitratesensor}, the designed system used WiFi-based IoT devices to monitor nitrate concentrations in ground water. In the design of the system, WiFi was selected for communicating the information due to its low cost, high throughput, and ease of integrating with web-based services. In~\cite{watersaving}, a WSN for irrigation control was developed using Zigbee for wireless communication. Zigbee was selected due to its low cost and off-the-shelf components to reduce the hardware complexity. In order to reduce the power consumption the transmit power was configured to be 0~dBm. In both of the systems presented, the power consumption was not a major concern, with a greater focus being placed on the total cost and ease of integrating with the system. If the network lifetime was a greater concern, more emphasis could be placed on reducing the power consumption.

In~\cite{energyManagement}, a study was performed on sensor nodes with solar energy harvesting capabilities. An energy management policy is used in order to produce the optimal throughput and minimizes the mean delay in the network. Another method of optimizing power consumption can be seen in~\cite{circuit}. Where a circuit was created for wireless sensor nodes where energy from a solar panel could be transferred to a rechargeable battery even if in poor weather conditions. 

In other works, it was determined that the sampling rate of the nodes can greatly affect the energy consumed and power supplied. In~\cite{samplingRate}, a method is presented to decide the sampling rate of sensor nodes to manage its energy more efficiently. Simulations demonstrated the efficiency of the proposed algorithm compared to other algorithms.

\begin{figure*}[t!]
\centering
  \includegraphics[width=0.8\textwidth]{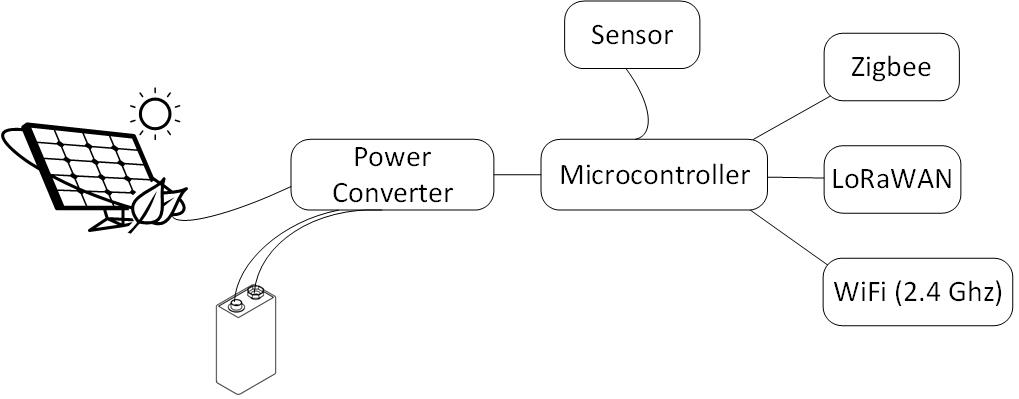}\\
  \caption{System framework used for experimentation.}\label{framework}
\end{figure*}

The system presented in~\cite{farmland} used a WSN in a cotton field to monitor soil moisture with automatic drip irrigation. Sensor nodes were developed to function using battery power while relay nodes functioned using solar power. A routing protocol was used in order to route data and increase power savings. Experimental results were conducted over a six month period and demonstrated that the system could function for a long period of time while collecting sensor information. 

Recently in literature, there have been many works that have used renewable energy sources for agricultural applications to extend network lifetime. In~\cite{hybrid}, due to the unpredictability associated with weather conditions, solar energy harvesting was combined with wireless charging in order to allow for nodes to function for longer periods of time. By combining the advantages of both solar energy harvesting and wireless charging it was found that based on experiments performed a significant increase in network performance could be achieved.

In this work, we expand on the papers described above and design a system for agricultural monitoring with energy harvesting capabilities. Based on the papers above there is a lack of research performed on using different types of wireless technologies for agricultural applications. In search of a system design, a comparison is performed between three wireless technologies: Zigbee, LoRaWAN, and WiFi to determine which technology consumes the lowest amount of power and provides the longest network lifetime. Three identical systems each using one of the wireless technologies to communicate are developed and tested. 

\section{System Framework} \label{frame}
The proposed system has a number of hardware components that were selected for measuring the power consumption of the different wireless technologies for agricultural applications. Each of the systems contained an Arduino Uno, a power converter, one rechargeable battery, a solar panel, a soil moisture sensor, and a communication unit. Each node forwards the sensor data to the base station using a different wireless technology. For Zigbee, a Series 2 2mW Wire Antenna XBee was used, for LoRaWAN, a Dragino LoRa Shield, and for WiFi, a CC3000 WiFi Shield.  The system framework is shown in Fig.~\ref{framework}.

\subsection{Components}
\begin{itemize}
\item \textit{Solar Panel:} To provide energy harvesting capabilities to the system, a Star Solar D165X165 monocrystalline solar panel was used~\cite{solarpanel}. Being only 170 x 170 x 2~mm, the solar panel is capable of providing a 6.0~V output at a peak of 3.65W when full sunlight is present. The small size makes it suitable for placement in a field where it would have minimal interference to any of the growing plants surrounding it while still providing a significant energy output.  The solar panel is shown in Fig.~\ref{solarpanel}.

\item \textit{Grand-Pro Lithium Polymer Battery:} The battery used was a Grand-Pro 3.7~V 6600~mAh Lithium Polymer (LiPo)~\cite{battery}. When connected to a battery the power converter was designed to provide a constant 5~V power output while the charge on the battery was above 3.4~V. If the charge dropped below 3.4~V the power converter would cease to function and would wait until the battery was sufficiently charged before supplying power again. This safety feature allowed the battery to maintain a voltage level preventing it from over-discharging and damaging the battery cells. The battery is shown in Fig.~\ref{battery}.

\item \textit{Power Converter:} A developed power converter was used to supply power to the sensor node systems~\cite{powerconverter}. In addition to supplying power, the converter was also capable of interfacing with other types of components such as an energy harvesting device and a recharge a Lithium Polymer (LiPo) battery. The energy harvesting device could then be used to both supply power to the node and recharge the battery. If it was no longer providing power the battery could then supply power to the system. The power converter is shown in Fig.~\ref{powerconv}.

\begin{figure*}[t!]
\centering
\subfloat[Solar Panel.]
{\includegraphics[width=.25\textwidth]{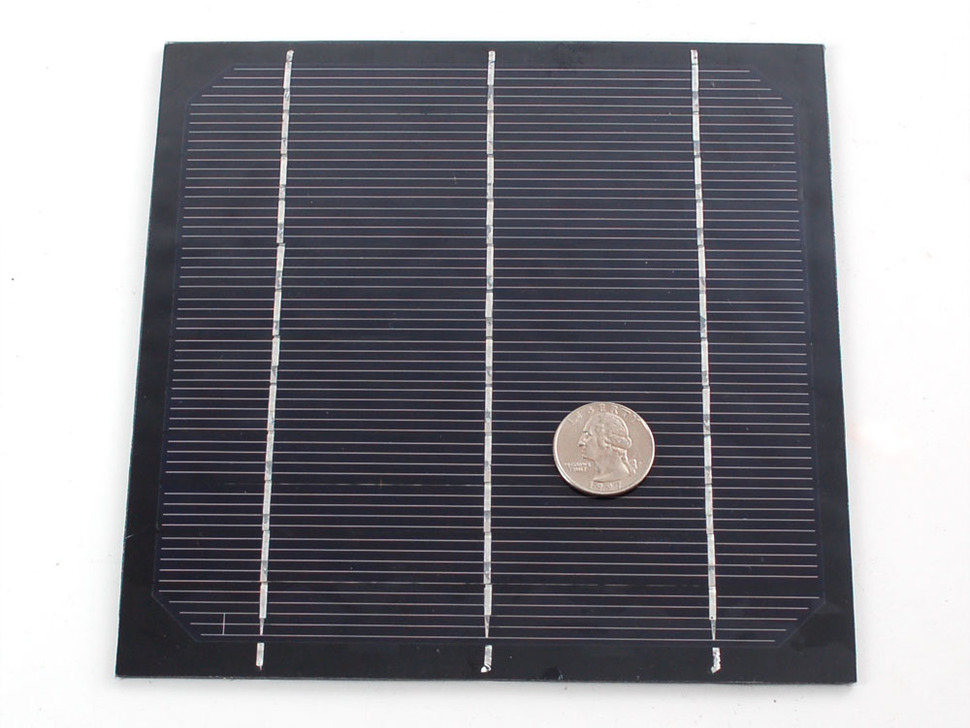}\label{solarpanel}} 
\subfloat[Battery.]
{\includegraphics[width=.25\textwidth]{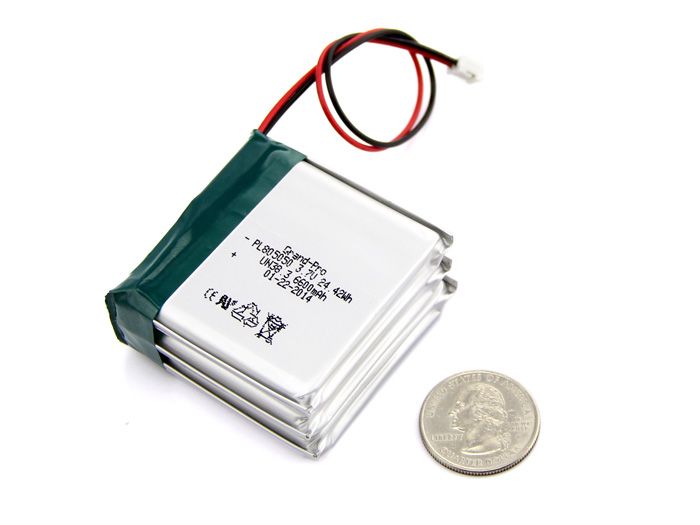}\label{battery}}
\subfloat[Power Converter.]
{\includegraphics[width=.20\textwidth]{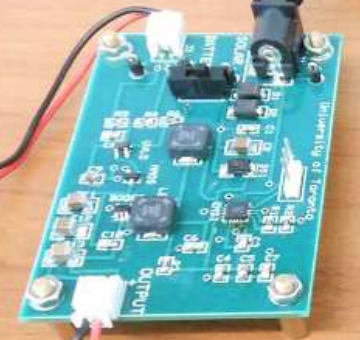}\label{powerconv}} 
\subfloat[Arduino Uno.]
{\includegraphics[width=.25\textwidth]{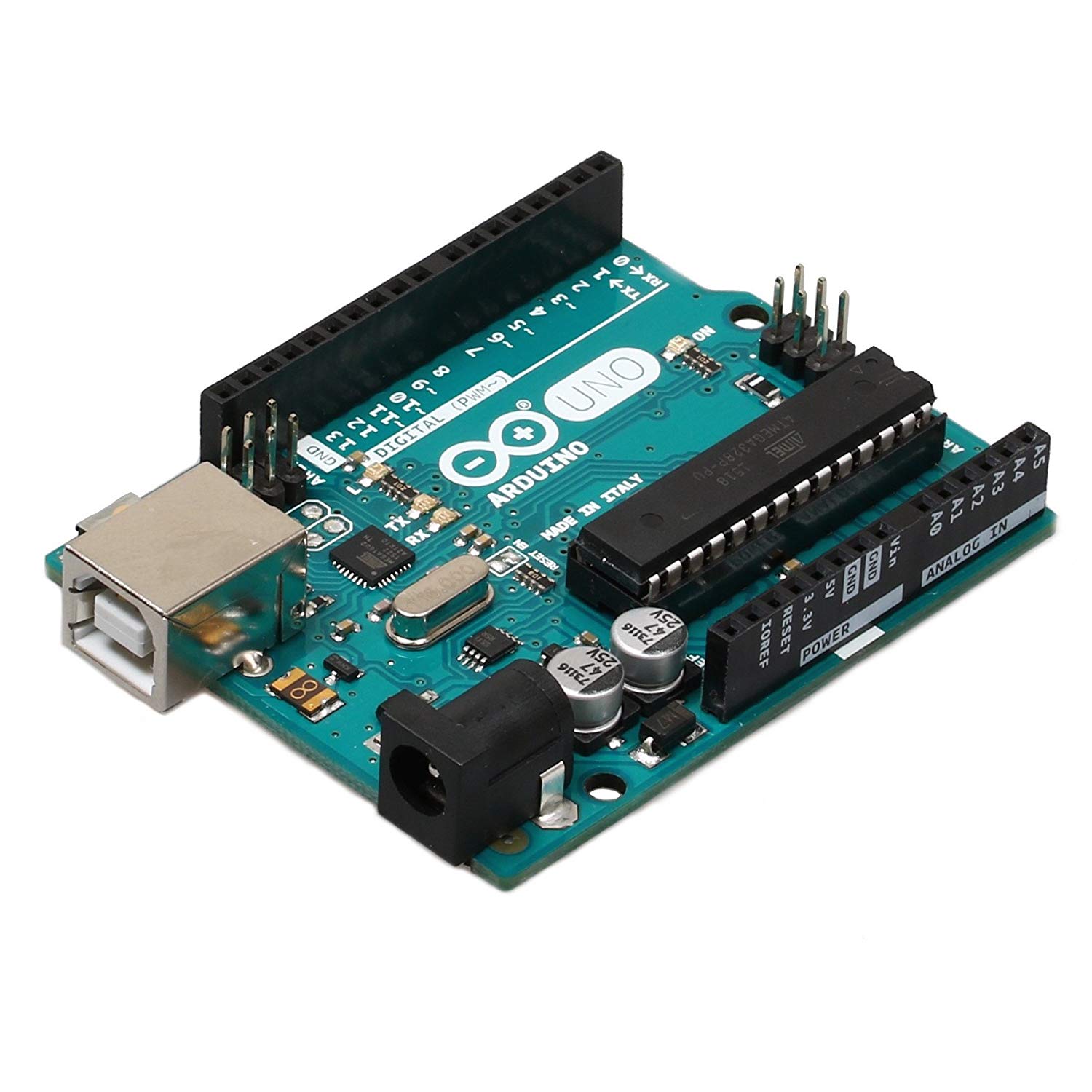}\label{arduino}} 
 
\subfloat[Soil moisture sensor.]
{\includegraphics[width=.25\textwidth]{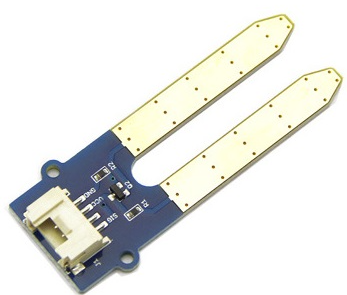}\label{sensor}} 
\subfloat[Series 2 XBee.]
{\includegraphics[width=.25\textwidth]{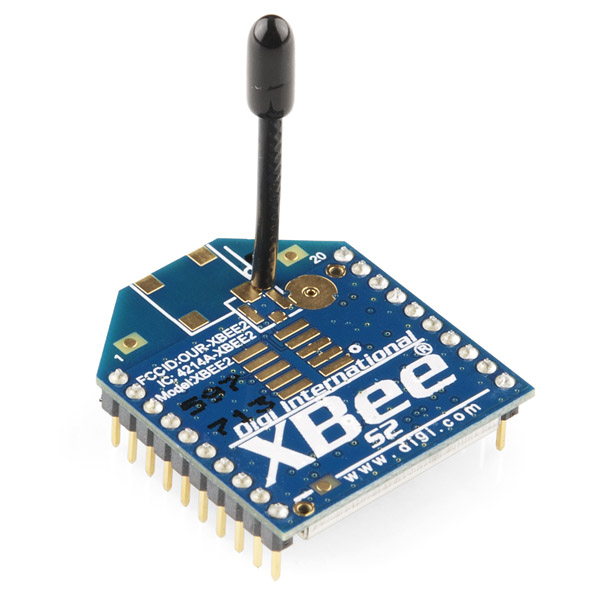}\label{xbee}} 
\subfloat[Dragino Lora.]
{\includegraphics[width=.25\textwidth]{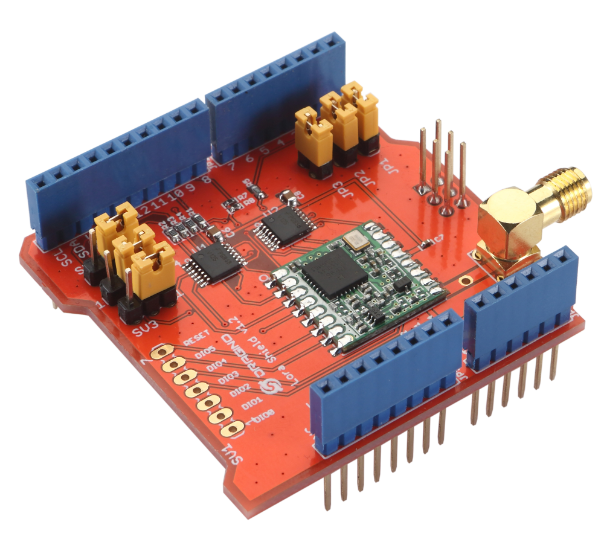}\label{lorashield}} 
\subfloat[WiFi (2.4Ghz).]
{\includegraphics[width=.25\textwidth]{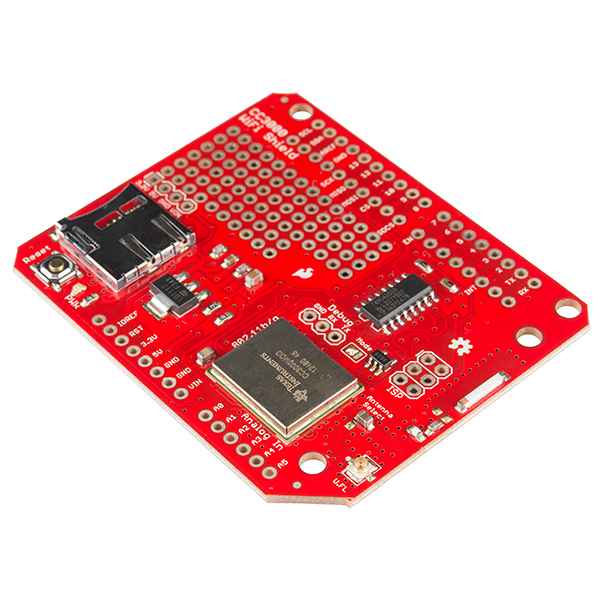}\label{wifishield}} 
\caption{Hardware components used for experimentation.}
\label{fig:test1}
\end{figure*}

\item \textit{Arduino Uno:} In order to connect all the hardware components together, an Arduino Uno Rev3 microcontroller was chosen~\cite{arduino}. The Arduino Uno was selected based on its low power consumption and ease of development in configuring all the components together. It is based on the ATmega328P, which contains six analog-to-digital converts that can be used to easily connect and read data from analog sensors. The Arduino Uno is shown in Fig.~\ref{arduino}.

\item \textit{Grove Soil Moisture Sensor:} To measure the moisture levels in the soil a Grove Soil Moisture Sensor was used~\cite{mosituresensor}. This sensor was selected as it draws a significant amount of current of reducing the device lifetime when environmental conditions are being measured. Soil moisture is also a commonly measured parameter in agricultural monitoring allowing for a system design similar to what would be used in a real-life application. The soil moisture sensor is shown in Fig.~\ref{sensor}.

\item \textit{Series 2 XBee with 2mW Wire Antenna:} To create a Zigbee network between the devices a pair of Series 2 XBees with a 2mW Wire Antennas were used~\cite{xbee}. The Series 2 XBees are low power radios which communicate on the Zigbee mesh network. These devices are capable of creating point-to-point or multi-point networks connecting together hundreds of nodes. Devices using Zigbee have a transmission range of 120~m in Line-of-Sight (LoS), which can provide many benefits for use in an agricultural monitoring system such as reducing the system costs and allowing for easy configuration of the devices. The Series 2 XBee  is shown in Fig.~\ref{xbee}.

\item \textit{Dragino LoRa Shield:} To communicate devices using LoRaWAN a pair of LoRa Shields for Arduino developed by Dragino were used~\cite{lora}. LoRaWAN is known for being a long range technology communicating at a low frequency of 915~MHz, signals produced have larger wavelengths hence can travel further distances. In LoS LoRaWAN is capable of transmitting up to 15000~m. Due to this, LoRaWAN is considered one of the best technologies to use for agricultural monitoring. Its large transmission range can greatly reduce the number of nodes required and its low power consumption can keep nodes functioning for a longer period of time compared to more commonly used technologies. The LoRa Shield  is shown in Fig.~\ref{lorashield}.

\item \textit{CC3000 WiFi Shield:} To connect the Arduino using WiFi a Sparkfun CC3000 WiFi Shield was used~\cite{wifi}. WiFi is one of the most commonly used wireless technologies, available in most devices, used to connect to a Wireless Local Area Network (WLAN) and the Internet. For agricultural purposes, WiFi is rarely used in the transmitting of information. WiFi has a short transmission range in LoS only capable of reaching up to 50~m distance. In addition, WiFi has a very large power consumption which often makes it a poor choice to use in wireless devices outdoors that require a power supply. The CC3000 communicates using the IEEE 802.11g standard. The WiFi shield is shown in Fig.~\ref{wifishield}.

\end{itemize}

\begin{table*}[t!]
\centering
\small
\caption{Summary of wireless technologies.}
\label{summary}
\begin{tabular}{|l|l|m{1.7cm}|m{1.7cm}|l|l|}
\hline
\textbf{Technology} & \textbf{Throughput} & \textbf{Transmission Range} & \textbf{Power Consumption} & \textbf{Advantages} & \textbf{Disadvantages} \\ \hline
Zigbee & 250 kbit/s & 120 m & Low & Easy to set up & Require extra hardware \\ \hline
LoRaWAN & 50 kbit/s & 15000 m & Extremely Low & Wide range & Requires extra hardware \\ \hline
WiFi (2.4Ghz) & 54 Mbit/s & 50 m & Moderate & Wide availability & High energy consumption \\ \hline
\end{tabular}
\end{table*}

\subsection{System Parameters}
When comparing different types of wireless technologies, parameters such as  transmission range and current consumption are important in determining the optimal technology for agricultural monitoring.  While the current consumption of a wireless technology is important in determining the longevity of a node's power supply, often other parameters should be considered first in the selection of a communication technology. Transmission range is one parameter that is often compared. By using devices that transmit further, a fewer number of relay nodes are required in order for a transmission to reach the intended destination. Another common parameter is the throughput. If a higher throughput is used, a larger amount of data can be transmitted in a period of time. Other parameters often used are the cost and ease of implementation, having a lower cost per-device can allow for a greater number of nodes to be implemented in the network. While having a simpler ease-of-implementation can allow for the network to have a lower set up time and make it easier to debug if a problem occurs.

\begin{table}[t!]
\renewcommand{\arraystretch}{1.2}
\centering
    \small

\caption{System parameters corresponding to components used in sensor nodes.}
\label{tb:parameters}
\begin{tabular}{|m{5.5cm}|r|}
\hline
\textbf{Parameter}				& \textbf{Value}	\\ \hline
Battery Current Supply            & 6600~mAh \\ \hline
Arduino Uno (Max Current Consumption)       & 45~mA   \\ \hline
Grove Soil Moisture Sensor (Max Current Consumption) & 35~mA \\ \hline
Series 2 XBee (Max Current Consumption)      & 40~mA   \\ \hline
Dragino LoRa Shield (Max Current Consumption)          & 10~mA   \\ \hline
CC3000 WiFi Shield (Max Current Consumption)          & 190~mA   \\ \hline
Sampling Frequency              & 1~Hz  \\ \hline
Tranmission Interval            & 1~s     \\ \hline
Transmission Power            & -10~dBm     \\ \hline
\end{tabular}
\end{table}

A summary of the parameters of the wireless technologies being compared in this paper can be seen in Table~\ref{summary}. In terms of throughput, WiFi can transmit the most amount of data reaching speeds of 54~Mbit/s. Zigbee is the next highest with 250~kbit/s, followed by LoRaWAN with 50~kbit/s. For transmission range, LoRaWAN is the optimal capable of reaching up to 15000~m in LoS. Zigbee was the second furthest in LoS with 120~m, while WiFi has the lowest transmission range only capable of reaching 50~m in LoS.

Other parameters such as the current consumption, the sampling frequency, transmission interval, and transmission power are important in the measuring of the power consumption of a device. Table~\ref{tb:parameters}, summarizes the current supplied by the battery, maximum current utilized by the various components, and the other parameters configured in the experiment that affect the power consumption, such as the sampling frequency, the  transmission interval, and the transmission power. 

In addition to the current draw of the components, while greatly affecting the power consumption, the transmission ranges of the wireless technologies are important when designing a system for agricultural monitoring. If devices are used that have a further transmission range, then a fewer number of nodes can be used in the monitoring of a field. A monitoring node would require an Arduino Uno, a sensor, a battery, and a communication unit. Table~\ref{node} shows the expected lifetime of each monitoring node when each wireless technology is used with maximum current consumption.

\begin{table}[t!]
\centering
\small
\caption{Estimated monitoring node max current consumption and min lifetime for each wireless technology.}
\label{node}
\begin{tabular}{|c|m{2.5cm}|c|}
\hline
\textbf{\begin{tabular}[c]{@{}c@{}}Wireless \\ Technology\end{tabular}} &  \textbf{\begin{tabular}[c]{@{}c@{}}Estimated Max \\ Current \\ Consumption (mA)\end{tabular}} & \textbf{\begin{tabular}[c]{@{}c@{}}Estimated \\ Min\\ Lifetime (h)\end{tabular}} \\ \hline
Zigbee-based node & 120 & 55 \\ \hline
LoRaWAN-based node & 90 & 73 \\ \hline
WiFi-based node & 270 & 24 \\ \hline
\end{tabular}
\end{table}

At the same time, if the communication units are used as relay nodes  to forward the data, the larger the transmission range the smaller the number of the required units to reach the destination. To achieve the maximum transmission range, the maximum current consumption is required. Table~\ref{antenna} shows the estimated lifetime, when each of the communication units is used as a repeater alone, to forward the sensor data to the destination, using the battery.

\begin{table}[t!]
 \centering
 \small
 \caption{Estimated relay node current consumption and lifetime for each wireless technology.}
 \label{antenna}
 \begin{tabular}{|c|c|c|}
 \hline
 \textbf{\begin{tabular}[c]{@{}c@{}}Wireless \\ Technology\end{tabular}} &  \textbf{\begin{tabular}[c]{@{}c@{}}Estimated Current \\ Consumption (mA)\end{tabular}} & \textbf{\begin{tabular}[c]{@{}c@{}}Estimated \\ Lifetime (h)\end{tabular}} \\ \hline
 Zigbee-based node & 40 & 165 \\ \hline
 LoRaWAN-based node & 10 & 660 \\ \hline
 WiFi-based node & 190 & 35 \\ \hline
 \end{tabular}
 \end{table}

\section{Experimental Procedure} \label{setup}
\begin{figure}[t!]
\centering
  \includegraphics[width=\columnwidth]{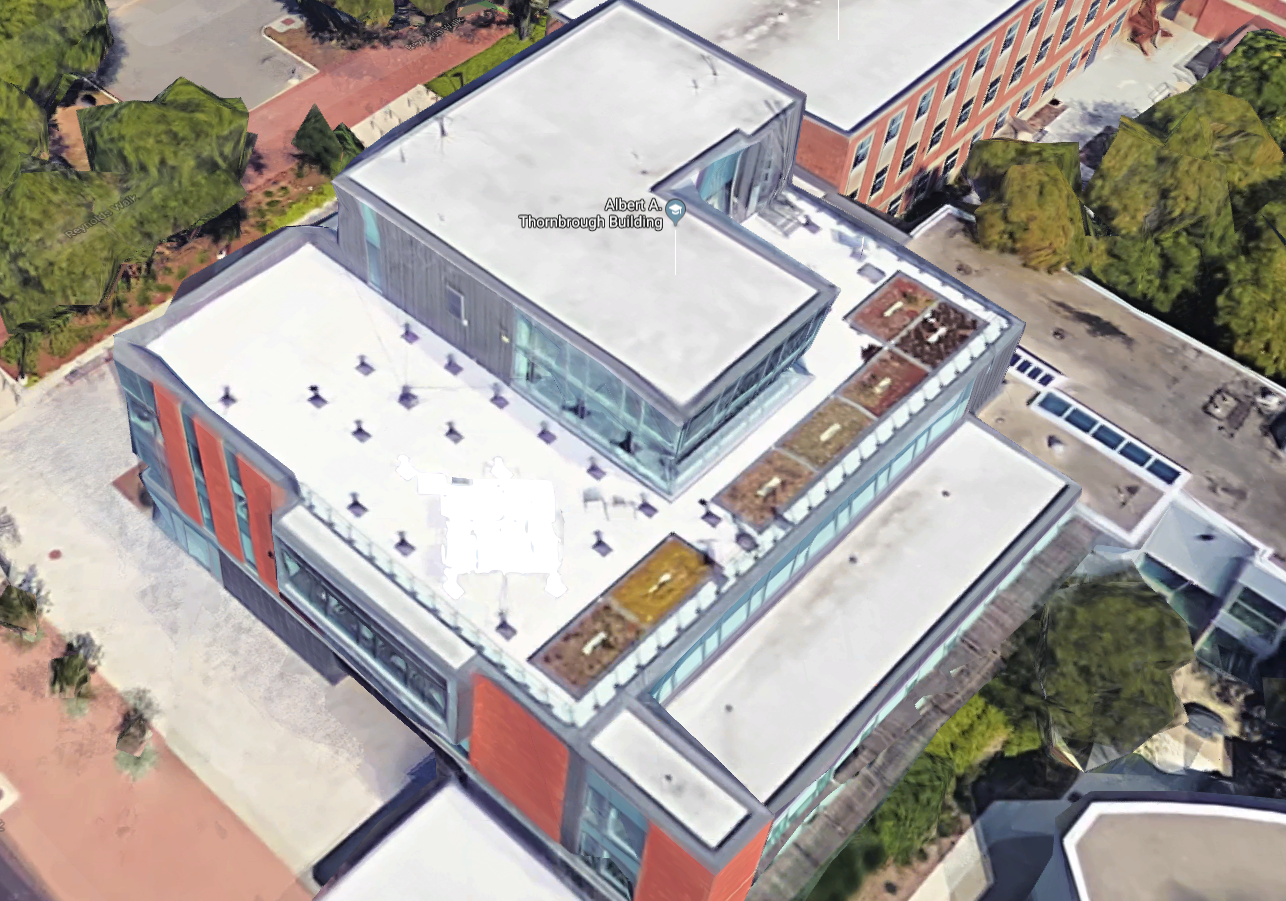}\\
  \caption{Green roof lab at University of Guelph.}\label{greenroof}
\end{figure}

In order to evaluate the proposed systems, a testbed was created to evaluate which wireless technology would be optimal for agricultural monitoring with energy harvesting. For each of the systems, identical nodes were configured except for their wireless communication method. The experiments were conducted at an outdoor environment where the solar panels for each of the nodes would obtain a similar amount of solar energy throughout the day. The testing area was a roof research lab at the University of Guelph Engineering Building, shown in Fig.~\ref{greenroof}. To measure the charge left on the battery, probes from the power converter were connected and measured on the Arduino and transmitted to a computer that was functioning as the destination. 

\subsection{Experimental Setup}

For testing purposes, nodes were configured to sample the charge left on the battery every 1~Hz and transmit the information every 1~s. Note that these times were used in order for the systems to consume a larger amount of power and therefore cease functioning sooner. If the systems were to be placed  in an actual environment for agricultural monitoring the times could be greatly reduced since actual conditions do not rapidly vary in a short period of time. Before starting the experiments, the batteries that were connected to the nodes were fully charged. 

\subsection{Outdoor Experiments}

Four experiments were performed, the first without energy harvesting capabilities and the following with energy harvesting capabilities. Each experiment lasted until all the nodes power supplies were drained and each of the nodes ceased to function:
\begin{enumerate}
\item \textit{Experiment 1 - No energy harvesting capabilities.} In this experiment, the solar power was not connected to each node, to examine and characterize the performance of the battery alone, without the solar panel.
\item \textit{Experiment 2 - With energy harvesting.} This experiment took during August 2018. 
\item \textit{Experiment 3 - With energy harvesting.} This experiment took during December 2018.
\item \textit{Experiment 4 - With energy harvesting.} This experiment took during May 2019.  
\end{enumerate}

Due to uncontrollable weather conditions performing three experiments would guarantee results that demonstrate the system performing with varying amounts of sunlight.

The current consumed by the nodes was also measured. In order to measure the current consumption of the devices, the Monsoon Power Monitor was used.  Monsoon is a monitoring tool that is capable of supplying an input voltage, measuring the current drawn by the device, and can display the average measurements. One useful function of the Monsoon is its ability to select a battery size and estimate the lifetime of the device based on that battery. To measure the current consumption of the devices, nodes were first powered and warmed up until the system was fully operational. The current was then measured for two minutes and the values recorded.

\begin{figure*}[t!]
\centering
\subfloat[Zigbee.]
{\includegraphics[width=.7\columnwidth]{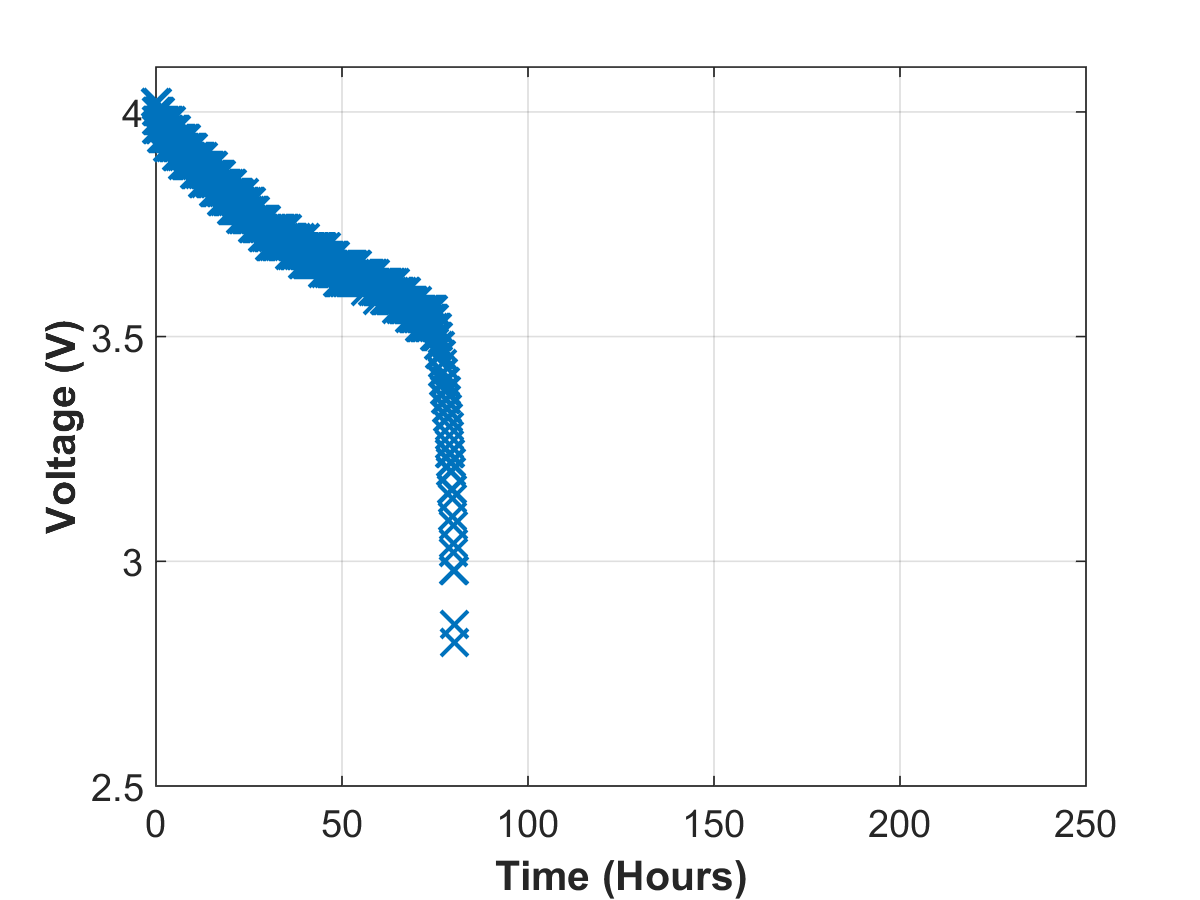}\label{zigbee0}} 
\subfloat[LoRaWAN.]
{\includegraphics[width=.7\columnwidth]{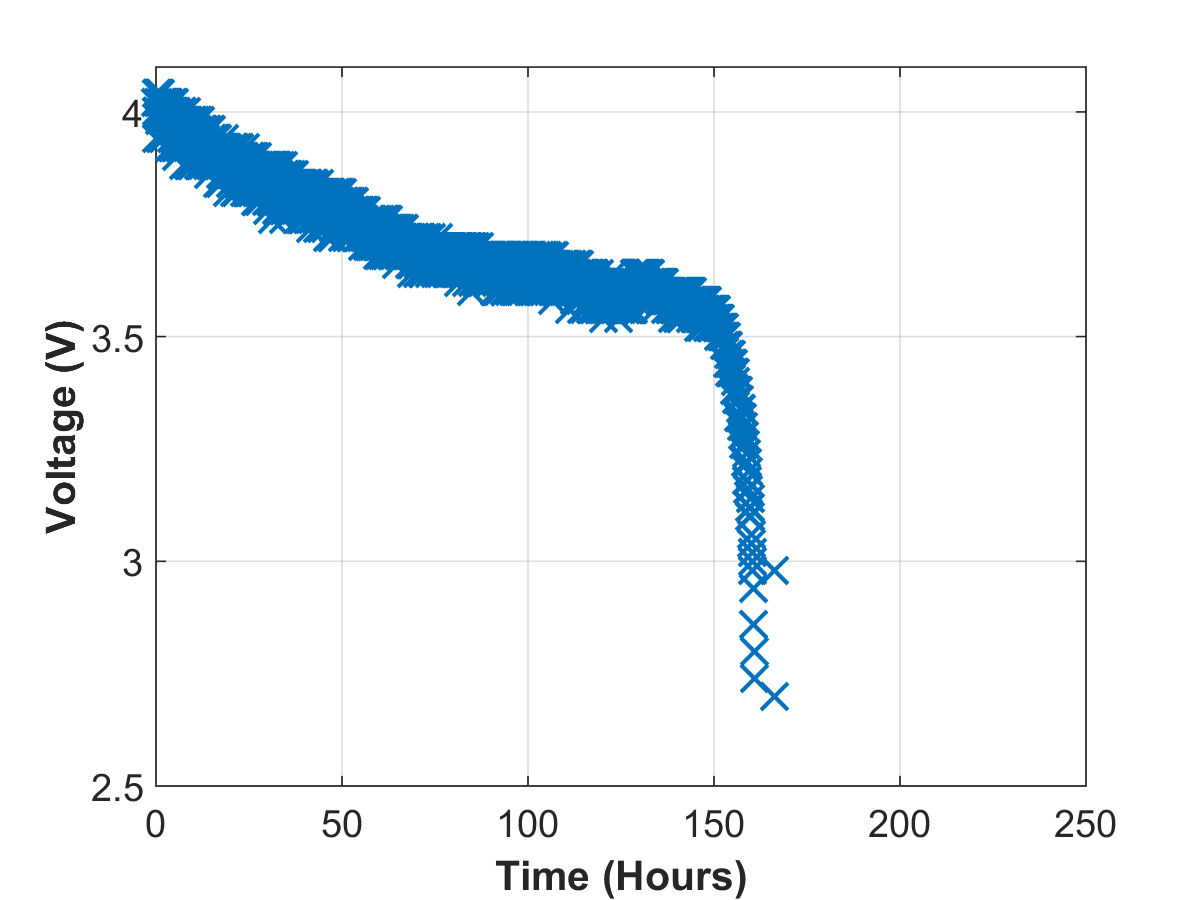}\label{lora0}} 
\subfloat[WiFi.]
{\includegraphics[width=.7\columnwidth]{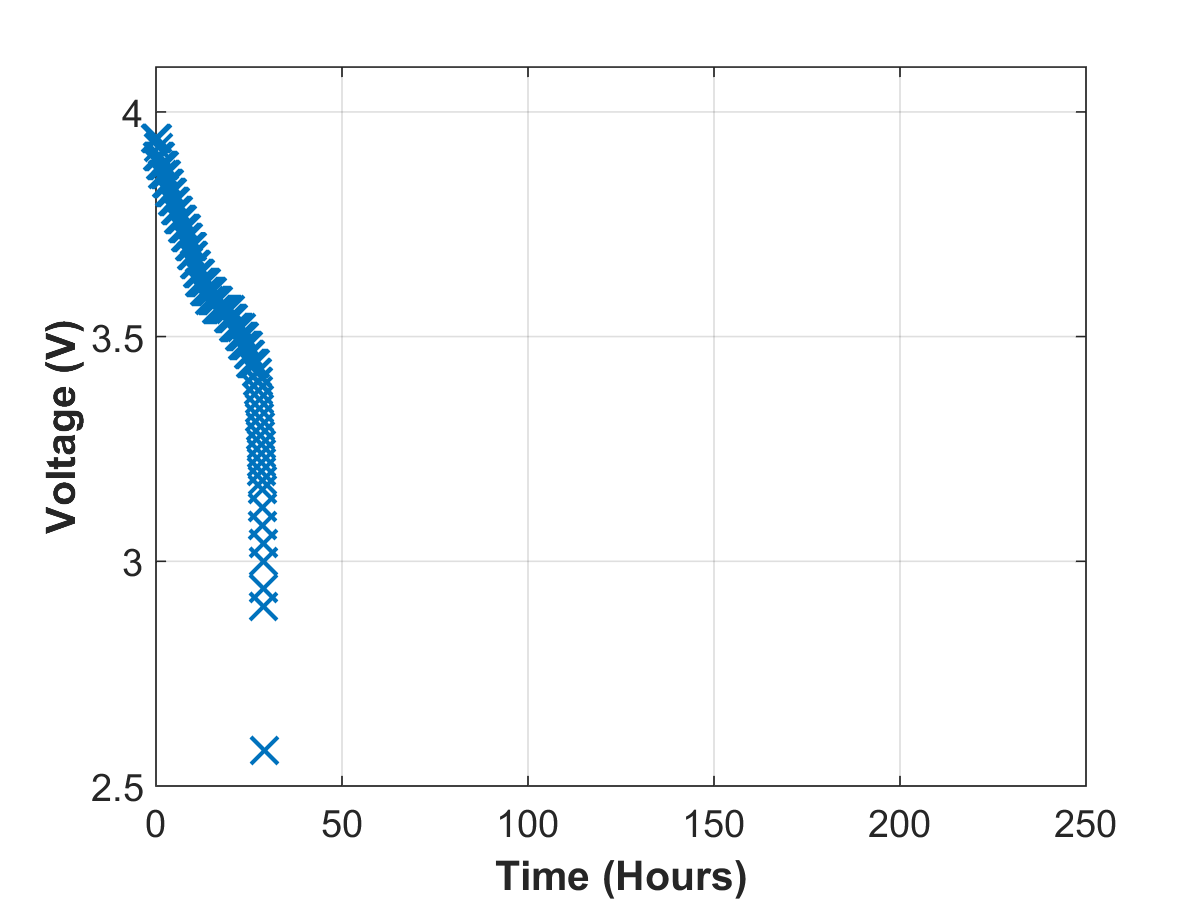}\label{wifi0}}
\caption{Experiment 1 - No energy harvesting capabilities.}
\label{fig:test1}
 \vspace{-3ex}
\end{figure*}

\begin{figure*}[t!]
\centering
\subfloat[Zigbee.]
{\includegraphics[width=.7\columnwidth]{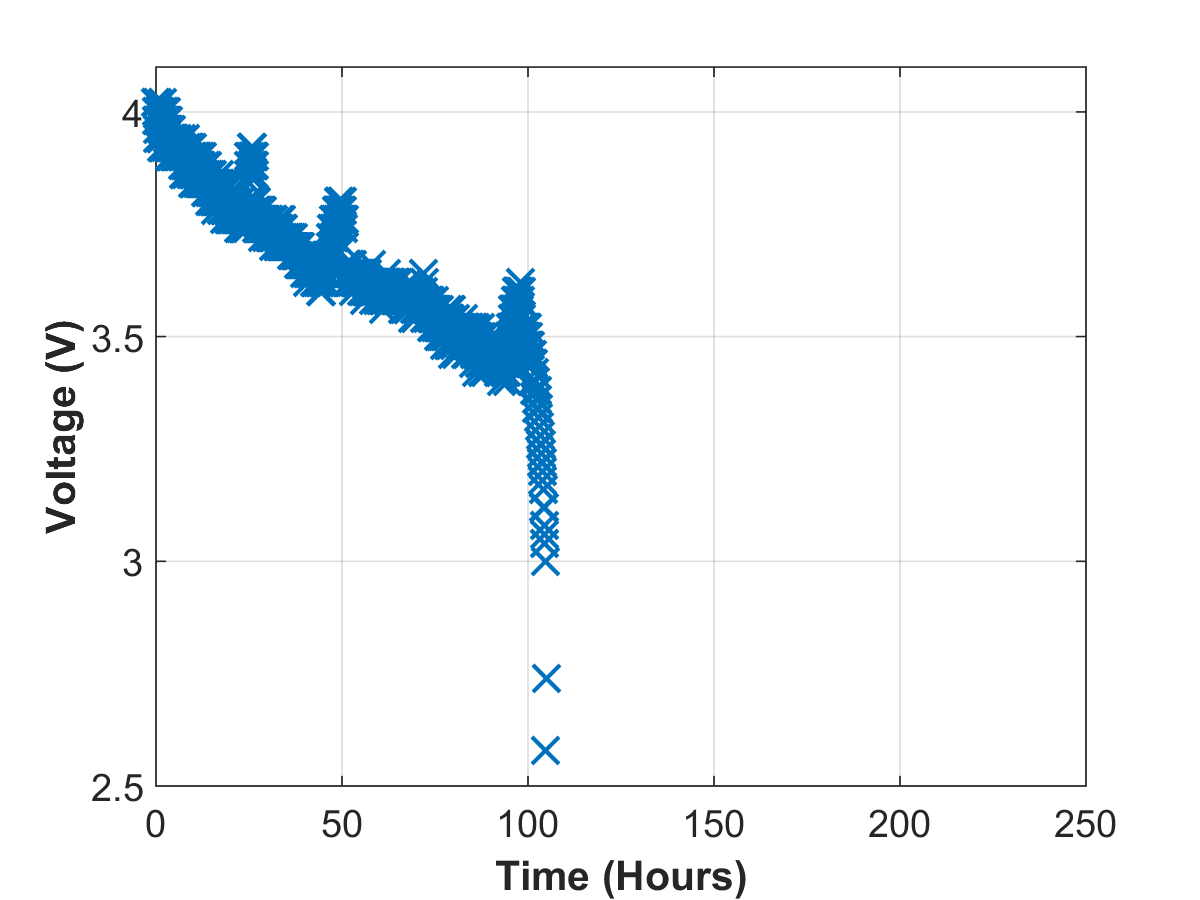}\label{zigbee1}} 
\subfloat[LoRaWAN.]
{\includegraphics[width=.7\columnwidth]{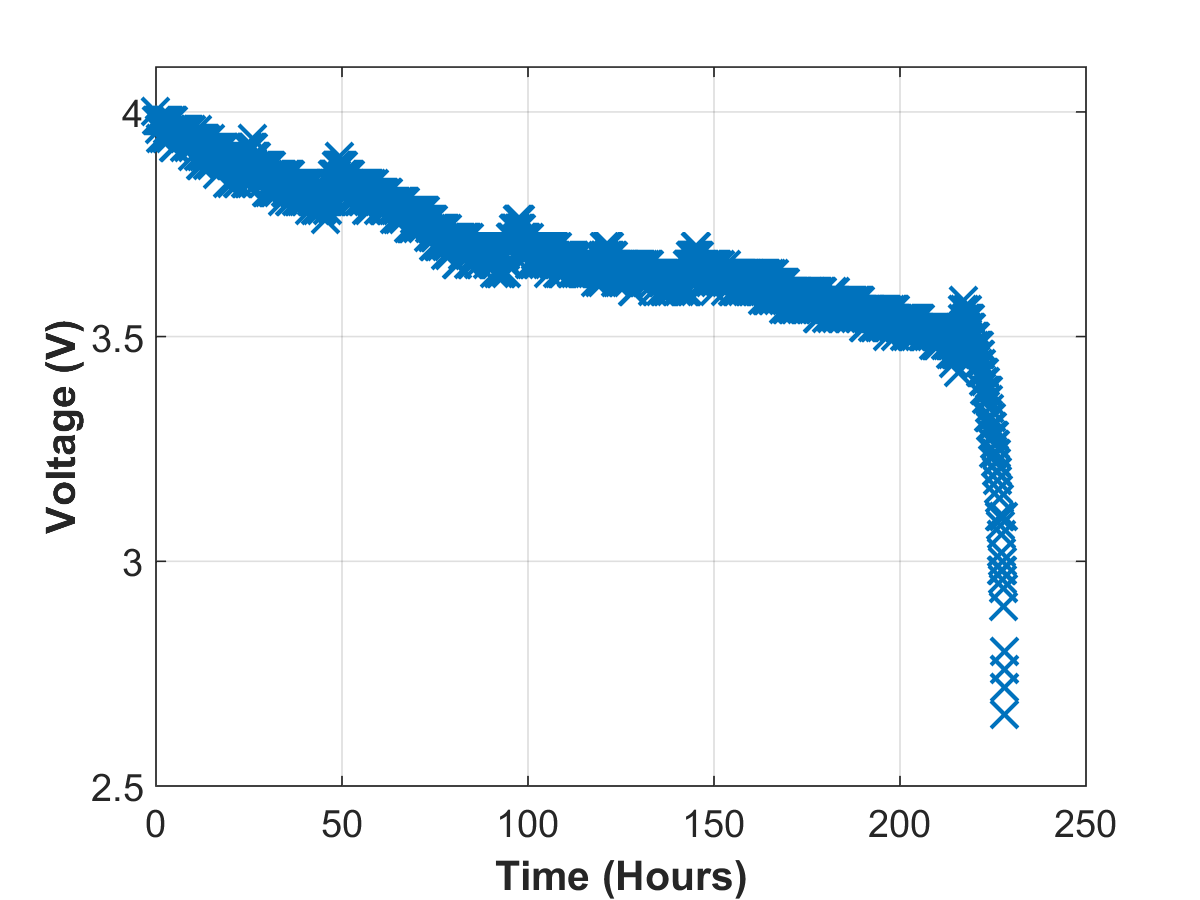}\label{lora1}} 
\subfloat[WiFi.]
{\includegraphics[width=.7\columnwidth]{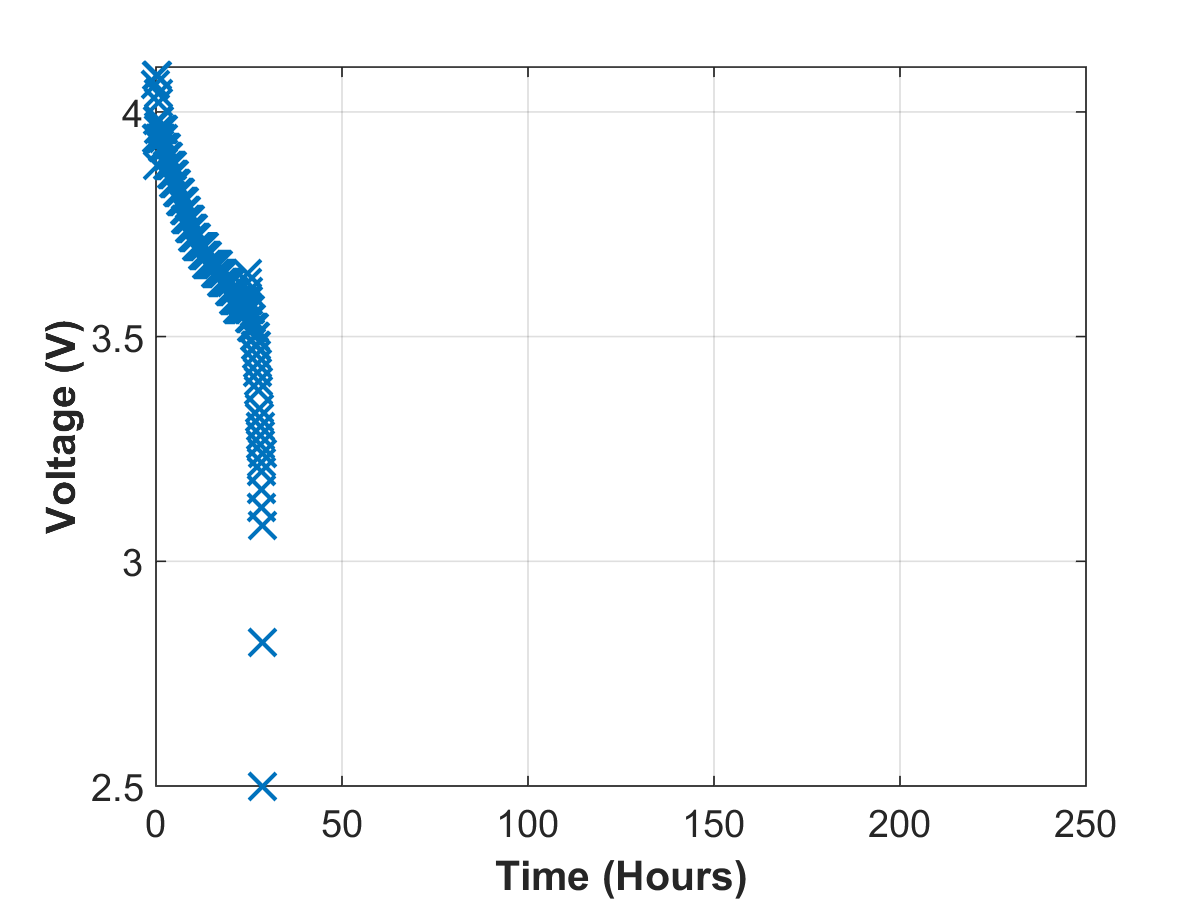}\label{wifi1}}
\caption{Experiment 2 - August 2018 with solar energy harvesting.}
\label{fig:test2}
 \vspace{-4ex}
\end{figure*}

\begin{figure*}[t!]
\centering
\subfloat[Zigbee.]
{\includegraphics[width=.7\columnwidth]{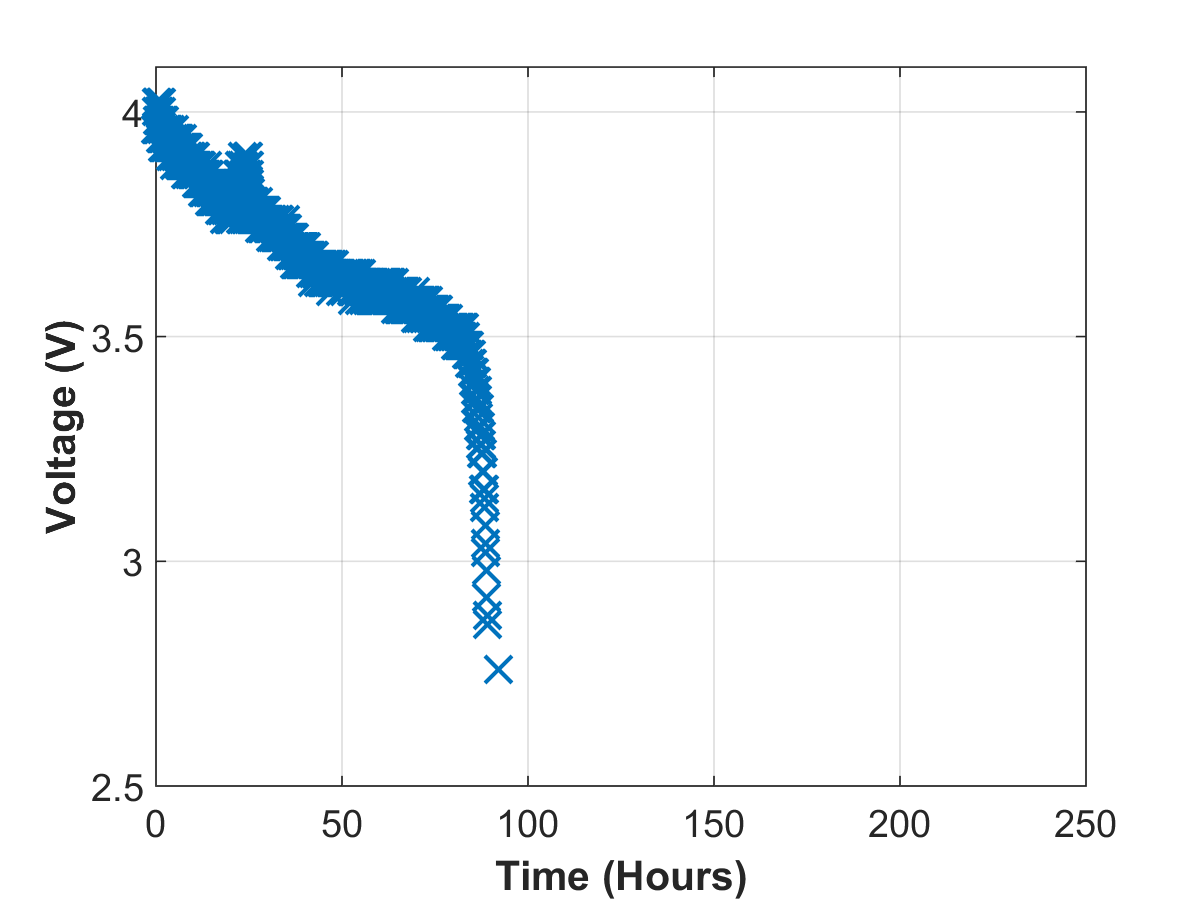}\label{zigbee2}} 
\subfloat[LoRaWAN.]
{\includegraphics[width=.7\columnwidth]{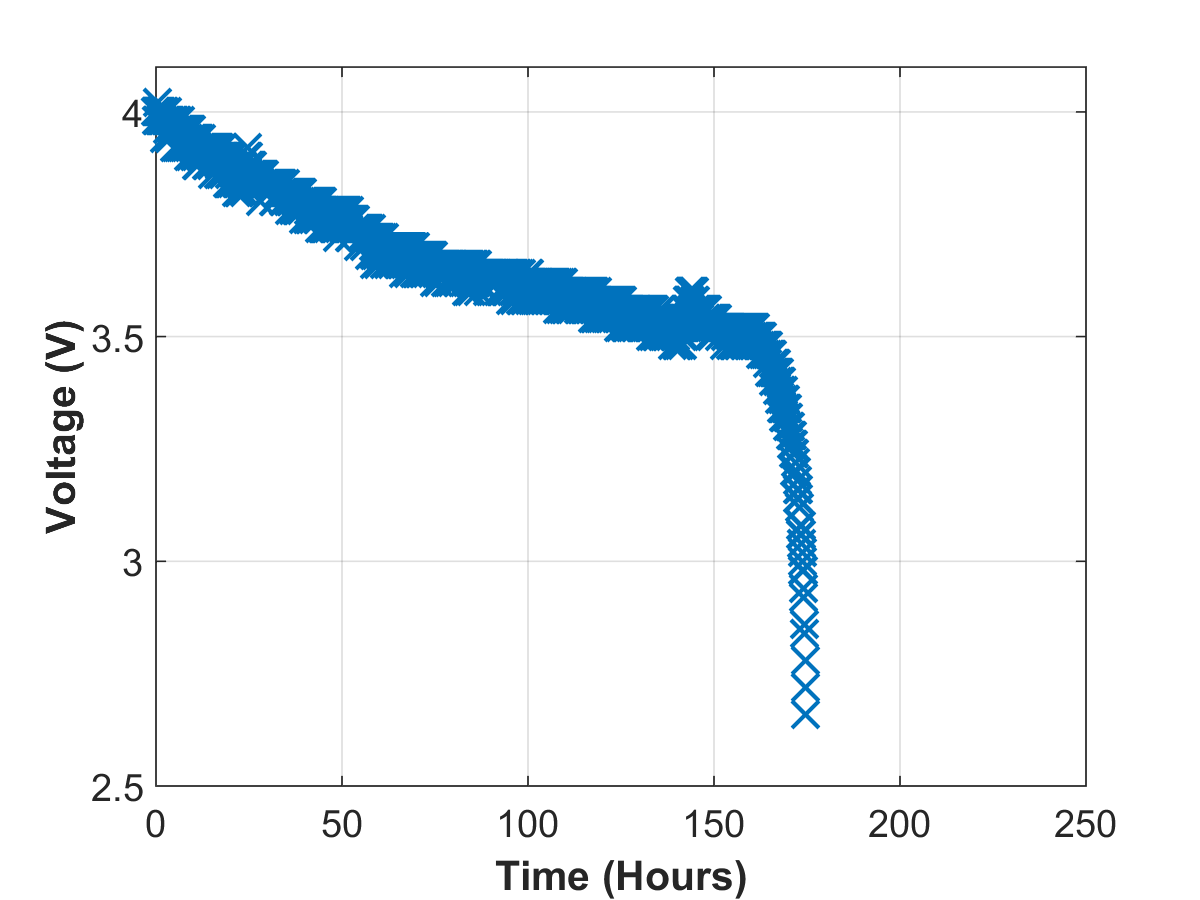}\label{lora2}} 
\subfloat[WiFi.]
{\includegraphics[width=.7\columnwidth]{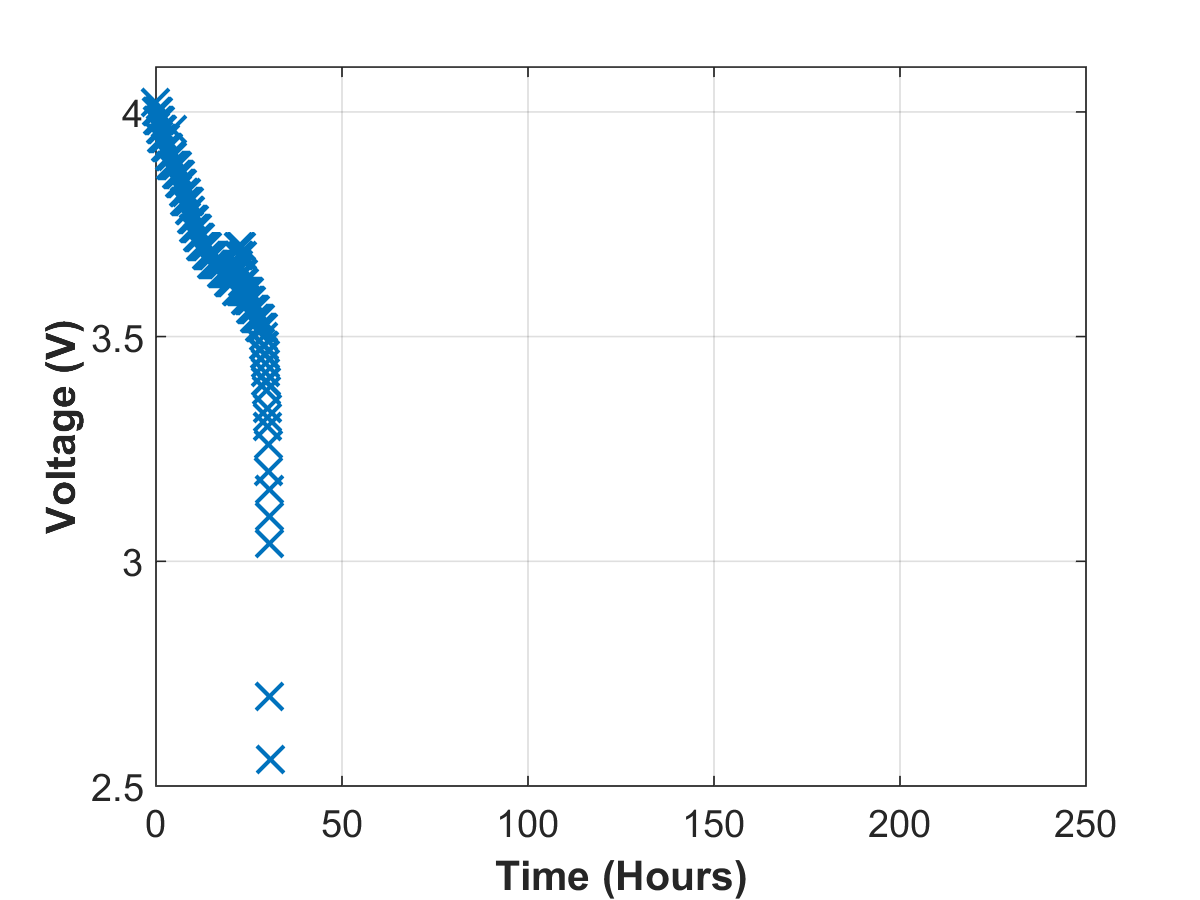}\label{wifi2}}
\caption{Experiment  3 - December 2018 with solar energy harvesting.}
\label{fig:test3}
 \vspace{-4ex}
\end{figure*}

\begin{figure*}[t!]
\centering
\subfloat[Zigbee.]
{\includegraphics[width=.7\columnwidth]{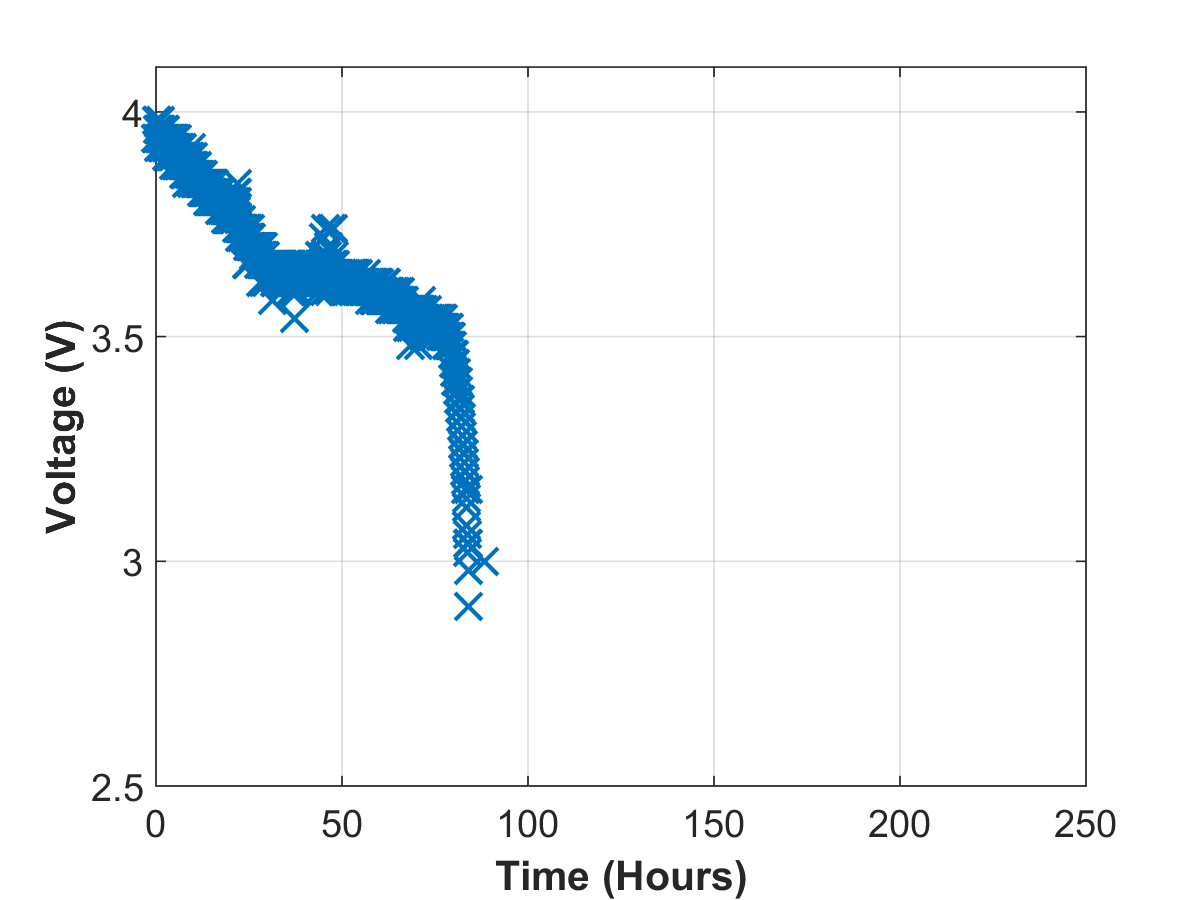}\label{zigbee3}} 
\subfloat[LoRaWAN.]
{\includegraphics[width=.7\columnwidth]{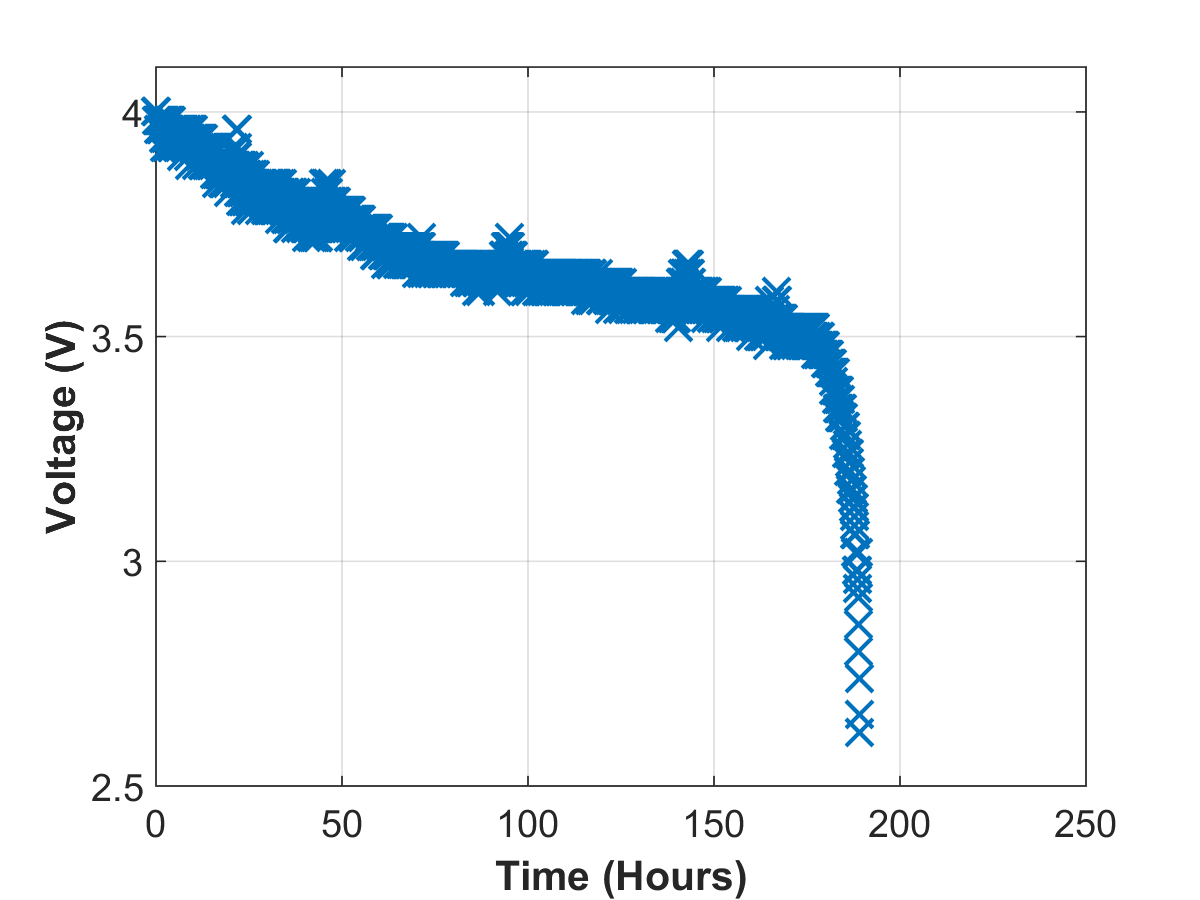}\label{lora3}} 
\subfloat[WiFi.]
{\includegraphics[width=.7\columnwidth]{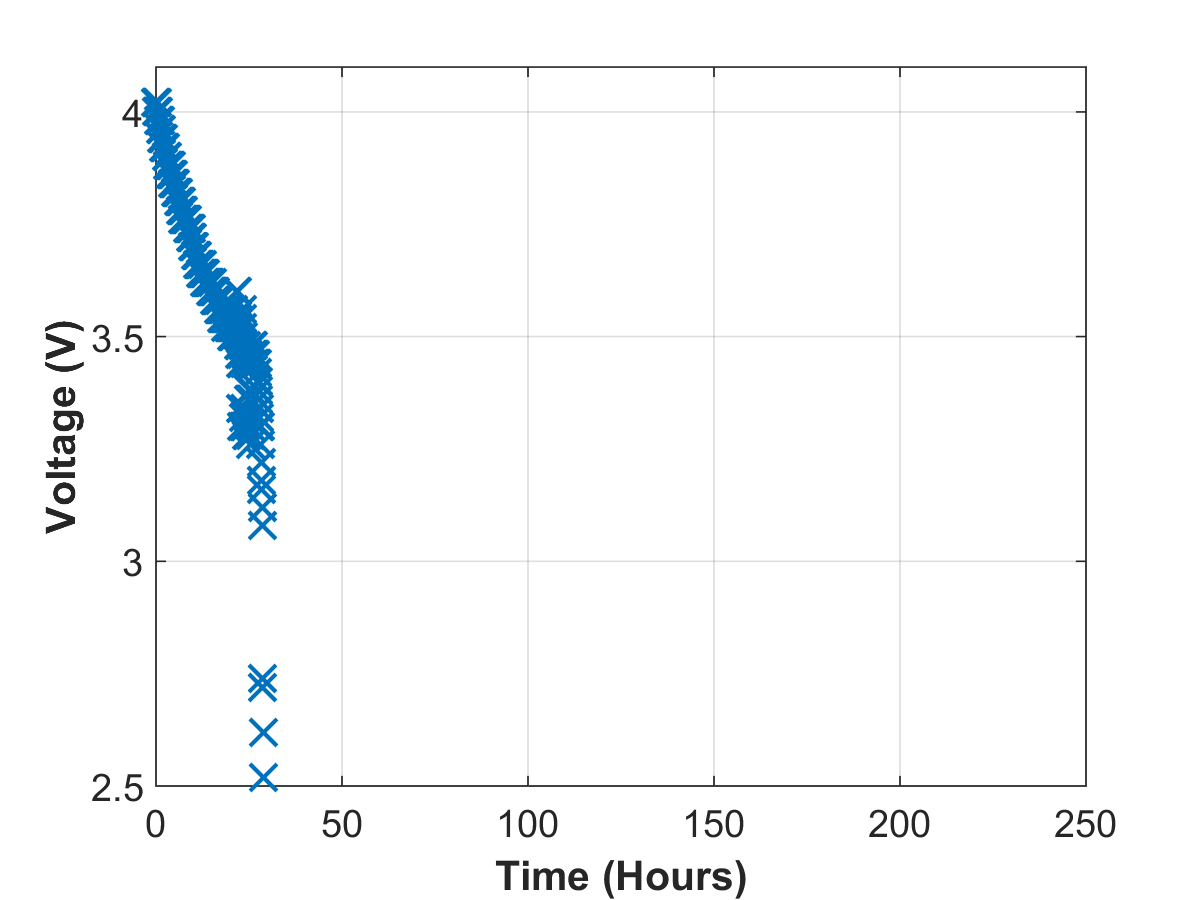}\label{wifi3}}
\caption{Experiment  4 - May 2019 with solar energy harvesting.}
\label{fig:test4}
 \vspace{-4ex}
\end{figure*}

\section{Results and Discussion} \label{results}
In this section, the experimental results are presented followed by a discussion on the acquired results.

\subsection{Results}
According to the experimental results obtained, each of the proposed systems functioned as required, capable of transmitting information until the battery in the sensor node was depleted. Due to the large amount of data that was gathered throughout the experiments, only a fraction was used and is displayed. 

The voltage charge remaining on the battery over time for the first experiment, with no energy harvesting capabilities, can be seen in Fig.~\ref{fig:test1}. 
and in Fig.~\ref{fig:test2}, Fig.~\ref{fig:test3}, and Fig.~\ref{fig:test4} the remaining battery levels over time for experiments 2, 3, and 4 is shown, respectively. An overall summary of the results gathered and the calculated measurements can be seen in Table~\ref{tb:summary}. 

Using the Monsoon Power Monitoring, the average current consumption's of the different devices along with the estimated lifetime could be measured. In terms of current consumption, it was determined that WiFi consumed the most requiring 171.17~mA to function. In second was Zigbee which required 69.36~mA and lastly, LoRaWAN was determined to use the lowest amount of current only consuming 29.33~mA on average. Using these values along with the battery size of 6600~mAh, the expected lifetime of the devices could be calculated. A Zigbee based system should function for 95.15~h, a LoRaWAN based system would be expected to run for 225.00~h, while a WiFi based system has an expected lifetime of 38.56~h. According to the experimental results, as seen in Fig.~\ref{fig:test1}, the Zigbee system functioned for 80.28~h before failing, LoRaWAN lasted for 166.23~h, and WiFi stopped after 29.06~h. 

 During the second experiment, shown in Fig.~\ref{fig:test2}, it was determined that the device using LoRaWAN technology was the most optimal capable of lasting 228.20~h on a single battery charge. The Zigbee based device was the second to cease operating stopping after 95.15~h. WiFi was determined to be the worst operating technology for transmitting data, only functioning for 38.56~h.

In the third experiment, shown in Fig.~\ref{fig:test3}, the results are similar to the previous experiments with fewer sunlight hours, forcing the nodes to use their batteries for power. This experiment took place during December 2018. The Zigbee based system was able to last 92.19~h, the LoRaWAN based system saw a large reduction in lifetime only functioning for 174.64~h, while the WiFi system experienced a similar runtime of 30.67~h.

Results from the fourth experiment can be seen in Fig.~\ref{fig:test4}. This experiment took place during May 2019 and  the solar panels manage to collect a greater amount of energy than the third experiment, but less than the amount gathered from the second experiment. In this case, the Zigbee system only functioned for 88.25~h. Using LoRaWAN the system was capable of running for 189.22~h, and the WiFi based system finished after 28.85~h.

\begin{table}[t!]
\centering
\caption{Summary of results.}
\label{tb:summary}
\begin{tabular}{|c|c|c|c|}
\hline
\textbf{\begin{tabular}[c]{@{}c@{}}Wireless \\ Technology\end{tabular}} & \textbf{Zigbee} & \textbf{LoRaWAN} & \textbf{WiFi} \\ \hline
\textbf{\begin{tabular}[c]{@{}c@{}}Average Current \\ Consumption (mA)\end{tabular}} & 69.36 & 29.33 & 171.17 \\ \hline \hline
\multicolumn{4}{|c|}{\textbf{No Energy Harvesting}} \\ \hline
\textbf{\begin{tabular}[c]{@{}c@{}}Expected \\ Lifetime (h)\end{tabular}} & 95.15 & 225.00 & 38.56 \\ \hline
\textbf{\begin{tabular}[c]{@{}c@{}}Experiment 1 - \\ Node Lifetime (h)\end{tabular}} & 80.28 & 166.23 & 29.06 \\ \hline \hline
\multicolumn{4}{|c|}{\textbf{Solar Energy Harvesting}} \\ \hline
\textbf{\begin{tabular}[c]{@{}c@{}} Experiment 2 - \\ Node Lifetime (h)\end{tabular}} & 104.80 & 228.20 & 28.3 \\ \hline
\textbf{\begin{tabular}[c]{@{}c@{}}Experiment 3  - \\ Node Lifetime (h)\end{tabular}} & 92.19 & 174.64 & 30.67 \\ \hline
\textbf{\begin{tabular}[c]{@{}c@{}}Experiment 4 -\\ Node Lifetime  (h)\end{tabular}} & 88.25 & 189.22 & 28.85 \\ \hline
\end{tabular}
\end{table}

\subsection{Discussion}
Through the experiments performed, it can be seen that LoRaWAN is the optimal technology for communicating information between nodes in a wireless network. By analysis the current consumption of the different devices, it could be determined that LoRaWAN consumed over a fifth the amount of current as WiFi and over half the current as Zigbee. Hence, this allows for LoRaWAN to last a much longer duration using batteries with similar capacities. The benefits LoRaWAN provides can be easily observed when a battery is selected to compare the lifetimes of the nodes. With a 6600~mAh battery, LoRaWAN is expected to last approximately 225.00~h, which is much greater than the 95.15~h Zigbee can provide, or the 38.56~h obtained from WiFi. 

When compared with the first experimental results, it can be determined that the estimated time calculated is not accurate. For instance, Zigbee was only capable of achieving a runtime of 80.28~h, while LoRaWAN functioned for only 166.23~h, with WiFi running for 29.06~h. Based on the real-world results, a great difference can be noticed between the two experiments. One reason for the difference can be attributed to the nonlinear discharge rate of the LiPo batteries. With all batteries not being ideal drops in the charge can occur reducing the lifetime of the device~\cite{battery1, battery2, battery3}. Another factor is the power converter. The power converter used to supply power to the device from the battery was designed to prevent the battery from over-discharging. Therefore, the power converter would stop the supplying power once the charge on the battery reached 3.4~V. 

In order to improve the battery life of the devices, the next set of experiments saw the addition of a solar panel to provide energy harvesting capabilities to the devices. Based on the results produced, adding energy harvesting to a system can greatly increase the lifetime of the nodes in the network. It was determined that the amount of sunlight obtained will greatly affect the additional lifetime that the device will be able to function for. The second experiment saw a large amount of sunlight supplying energy to the devices, with the third experiment supplying a very little amount of energy, and the fourth experiment providing energy to be between the previous two experiments.

In the Zigbee device, solar harvesting was able to greatly increase the lifetime of the node during the experiment. The second experiment saw Zigbee run for 104.80~h, 92.19~h during the third experiment, and 88.25~h in the fourth experiment. When compared to the first experiment, with energy harvesting a Zigbee node could last for an addition 25~h with a large amount of sunlight, while for a low amount lasted for 88.25~h. This is a great improvement for the Zigbee system as being able to function for a greater period of time would allow for more data to be gathered before the battery in the node would be to be recharged or replaced. 

The LoRaWAN based system, a larger amount of variance between the runtimes could be observed. The second experiment saw the node run for 228.20~h, the third for 174.64~h, and the fourth for 189.22~h. Due to the long base runtime of the device, it can be noticed that a larger amount of solar harvesting could occur further increasing the runtime of the device. At its peak, the second experiment saw the device last for an additional 50~h before failing, while in the third experiment the solar panel provided very little benefits to the system. 

Lastly, a WiFi based system using energy harvest has very little impact on the runtime of the device. During the second experiment the node runs for 28.3~h, while in the third experiment for 30.67~h, and for 28.85~h in the fourth experiment. Overall, the largest impact gained from energy harvest was approximately 1~h. A system using WiFi consumed too much power draining the battery charge that using energy harvesting no impact could be made on the system. 

In the WiFi experiments, it can be seen that the solar panel provided little benefits. There was a very little amount of battery charge recovered over the period that the node functioned. For the Zigbee and LoRaWAN systems, the solar panel provided much more energy and made a bigger difference in the system. It can clearly be seen the points when the solar panel was providing energy, after the solar panel stopped providing energy the charge on the battery was slightly increased.  

A number of other parameters such as the sampling frequency, transmission interval, and transmission power exist which could also affect the estimated and actual lifetime of the systems. In order to determine how much each of the parameters affect the power consumption additional experimentation would need to be performed. 

According to the results determined in Table~\ref{tb:summary}, WiFi is ideal if a large amount of information is required to be transmitted between short distances. However, at the cost of such a high speed, a greater amount of power consumed. On the other hand, LoRaWAN has a much lower throughput, but is able to transmit far distances with a very minimal amount of power being consumed. In the middle there is Zigbee. Zigbee has a slightly higher throughput than LoRaWAN, but a greatly reduced transmission range. The power consumed by Zigbee is still low and a network can be easily set up with nodes capable of being easily configurable and meshed together in the network.

\section{Conclusions} \label{con}
In this paper, we provide an experimental analysis between three wireless technologies: Zigbee, LoRaWAN, and WiFi when they are used in an agricultural monitoring system with energy harvesting capabilities. Identical systems were created each functioning with a wireless technology. The systems were placed outdoors and the batteries could be recharged. The systems were compared on the lifetime of the nodes, where the node that functioned for the longest time would be the most optimal for an agricultural application. Experimental results demonstrated that LoRaWAN would be ideal as it was capable of functioning for the longest period of time before failing. Zigbee was the next ideal, followed by WiFi. 

However, power consumption and device lifetime are usually not the only parameters that are considered when designing a system. While WiFi has a poor power consumption, the throughput is much higher allowing for a larger amount of information that can be transmitted between devices. The results produced in the paper can be used as an indicator for the selection of a wireless technology to be used in an agricultural monitor system with energy harvesting capabilities. 

\section*{References}

\bibliography{ref}
\end{document}